\documentclass
[preprint,10pt,twocolumn,prl,letterpaper,noshowpacs,balancelastpage]{revtex4}%
\usepackage{amsfonts}
\usepackage{amsmath}
\usepackage{amssymb}
\usepackage[dvips]{graphicx}
\usepackage{hyphenat}%
\providecommand{\U}[1]{\protect\rule{.1in}{.1in}}

\begin{document}
\preprint{ }
\title{Interaction induced spatial correlations in a Disordered Glass}
\author{Z. Ovadyahu}
\affiliation{Racah Institute of Physics, The Hebrew University, Jerusalem 9190401, Israel }


\begin{abstract}
A consequence of the disorder and Coulomb interaction competition is the
electron-glass phase observed in several Anderson-insulators. The disorder in
these systems, typically degenerate semiconductors, is stronger than the
interaction, more so the higher is the carrier-concentration \textit{N }of the
system. Here we report on a new feature observed in the electron-glass phase
of In$_{\text{x}}$O with the lowest \textit{N} yet studied. The feature,
resolved as a broad peak in field-effect measurements, has not been recognized
in previously studied Anderson-insulators. Several empirical facts associated
with the phenomenon are consistent with the conjecture that it reflects a
correlated charge-distribution. In particular, the feature may be turned on
and off by gate-voltage maneuvering, suggesting the relevance of
charge-arrangements. It may also be suppressed by either; temperature,
non-ohmic field, or exposure to infrared illumination. After being washed-out,
the feature reappears when the system is allowed to relax for sufficiently
long time. A puzzling aspect that arises is the apparent absence of the
phenomenon when the carrier-concentration increases above a certain value.
This is reminiscent of the glass-transition conundrum except that the role of
temperature in the latter is played by disorder. Analysis of these findings
highlights several issues that challenge our understanding of the
disorder-interaction interplay in Anderson insulators.

\end{abstract}
\maketitle

\section{Introduction}

The interplay between disorder and Coulomb interactions has been the subject
of intense research, mostly devoted to degenerate Fermi systems \cite{1,2,3,4}%
. A common feature observed in disordered-interacting condensed-matter systems
is a local depression in their single-particle density-of-states $\rho
$($\varepsilon$). This feature, referred to as a zero-bias anomaly (ZBA), is
anchored to the chemical potential of the system and appears in both, the
diffusive \cite{5,6}, and insulating regimes \cite{7,8}. A zero-bias anomaly
may be observed when inserting a particle into the many-body system faster
than the time it takes existing particles to relax to its presence. This may
be accomplished in a tunneling or photoemission measurement of disordered
metals and doped semiconductors. When the system is quantum-coherent, the
process is closely related to the Anderson orthogonality catastrophe \cite{9}.

An indirect way to monitor $\rho$($\varepsilon$), becomes possible in systems
where the disorder exceeds the critical value for Anderson-localization; the
competition between disorder and the unscreened Coulomb interaction slows down
the medium relaxation which, in turn, makes the system $\rho$($\varepsilon$)
observable in field-effect experiments \cite{10,11,12,13,14,15,16,17,18}. In
this case a ZBA appears as a modulation of the conductance versus
gate-voltage, G(V$_{\text{g}}$) \cite{14}. This feature, called a memory-dip
(MD), has been observed in several heavily-doped semiconductors where the
disorder necessary to render them Anderson-insulators is large enough to
reduce electronic relaxation rates many decades below the transition-times
associated with their conductivity \cite{19}. Anderson insulators in this
group exhibit glassy dynamics and are referred to as `electron-glasses'. These
should be distinguished from lightly-disordered systems (sometimes called
Coulomb-glasses) that do not exhibit MD in field-effect experiments \cite{19}.
The relation between the single-particle density-of-states and the MD has been
elucidated in \cite{12,14,17,18}.

Theoretical models that incorporate disorder and interactions are usually
concerned with the low-energy part of the single-particle density-of-states
The energy location of the states were expelled from the ZBA region has
received less attention. In particular, the Anderson-localized regime, where
$\rho$($\varepsilon$) has traditionally been derived via a classical
Coulomb-gap approach, the missing states are usually depicted as being evenly
spread outside the depleted region \cite{20,21,22,23}. Given that disorder in
Anderson insulators is in general significantly stronger than the Coulomb
interaction \cite{24}, this picture seems plausible and is consistent with
results of tunneling \cite{25,26}, and field-effect experiments on strongly
disordered systems \cite{27}.

Recently however, a strikingly different behavior was detected in a particular
version of amorphous indium-oxide In$_{\text{x}}$O films; as shown in figure 1.%

\begin{figure}[ptb]%
\centering
\includegraphics[
height=2.4785in,
width=3.3399in
]%
{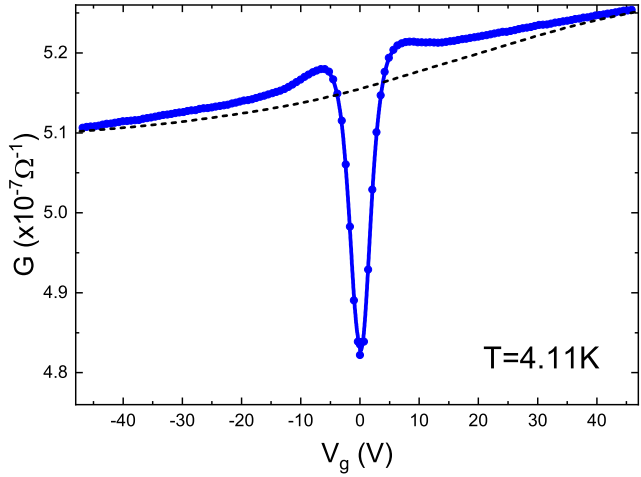}%
\caption{Conductance G vs. gate-voltage V$_{\text{g}}$ for a 20nm thick
In$_{\text{x}}$O film with carrier-concentration \textit{N}$\approx
$10$^{\text{19}}$cm$^{\text{-3}}$. The sample is separated from a degenerate
Si:B gate by 2$\mu$m layer of SiO$_{\text{2}}$. The figure shows a memory-dip
centered at V$_{\text{g}}$=0 where the sample was equilibrated for an hour
after being cooled from room-temperature. The dashed line delineates the
energy dependence of $\partial$n/$\partial$$\mu$, the material thermodynamic
density-of-states..}%
\end{figure}
The MD exhibited by specimen of this version reveal structure that resembles
that of a superconductor $\rho$($\varepsilon$) with its characteristic
`coherence-peaks' at the edges of the ZBA. The version of In$_{\text{x}}$O
used in this work is \textit{not} superconducting (at least down to $\approx
$0.28K) even when the system is in the diffusive transport regime \cite{28}.
Moreover, the indium-rich version of In$_{\text{x}}$O, which \textit{is}
superconducting when its disorder is sufficiently small, does show a
memory-dip when strongly-localized but without these side-shoulders \cite{27}.
Different versions of In$_{\text{x}}$O are distinguished by their In/O
composition that determines their carrier-concentration \textit{N} \cite{28}.
In terms of microstructure they are quite similar. Electron-diffraction
patterns of the superconducting version of the compound and that of the low
carrier-concentration sample used in this study are shown in Fig.2. A scan of
the radial intensity distribution of these patterns shown in Fig.3 is required
to be able to detect any difference between the two versions. Detailed study
of the In$_{\text{x}}$O versions including x-ray diffraction and
interferometry, and Raman-spectrometry showed that in terms of structural
properties, different versions only differ by small quantitative aspects like
the position of the boson-peak \cite{29}. However, in terms of disorder
perceived by the charge carriers, the difference may be substantial when
comparing In$_{\text{x}}$O with similar resistivities but different
carrier-concentrations. The system with the lower carrier-concentration has a
lower disorder (given the same resistivity), and therefore it may also be more homogeneous.%

\begin{figure}[ptb]%
\centering
\includegraphics[
height=1.8585in,
width=3.3399in
]%
{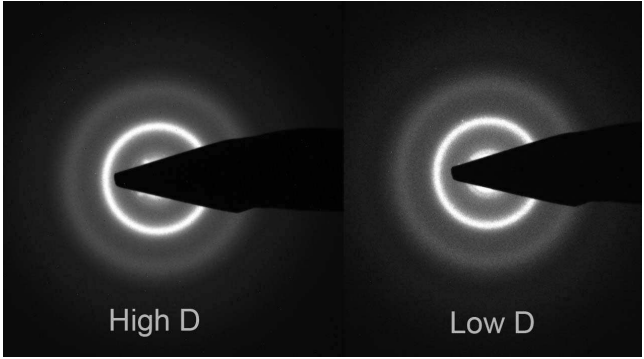}%
\caption{Electron diffraction-patterns of two versions of In$_{\text{x}}$O
films 20nm thick. The left pattern (marked High D) has carrier-concentration
\textit{N}$\approx$8x10$^{\text{20}}$cm$^{\text{-3 }}$ the right pattern
(marked Low D) is characteristic of the batch used in the present study with
\textit{N}$\approx$10$^{\text{19}}$cm$^{\text{-3 }}$}%
\end{figure}
%

\begin{figure}[ptb]%
\centering
\includegraphics[
height=2.3644in,
width=3.3399in
]%
{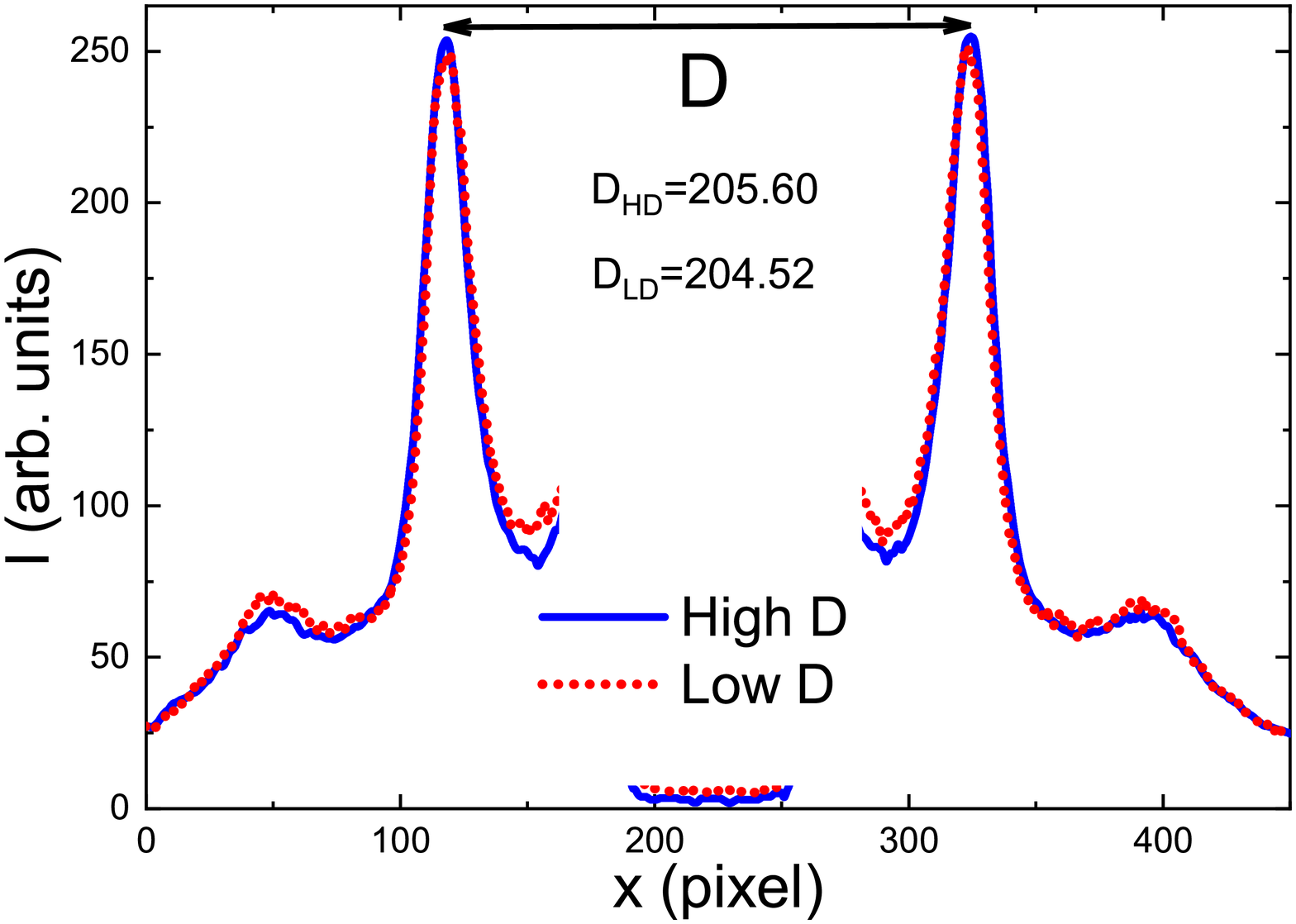}%
\caption{Intensity profile scan of the two diffraction-patterns shown in
Fig.2. The first ring diameters (in units of pixels) are shown for comparison.
Note the similarity and the small difference in the nearest-neighbor distance
of the two versions.}%
\end{figure}

Theoretically, piling-up of states at the edges of the ZBA is consistent with
the sum-rule for $\rho$($\varepsilon$). A non-monotonic $\rho$($\varepsilon$)
above the gap was demonstrated in numerical studies of disordered and
interacting quantum-systems, explicitly in strongly-correlated compounds
\cite{30,31,32,33,34} and in quantum-dots \cite{35}. A "shouldered" $\rho
$($\varepsilon$), quite similar in shape to the MD exhibited in Fig.1, was
obtained in Monte-Carlo simulations based on a random-displacement version of
the Coulomb-glass \cite{36,37,38}. The shoulders in this scenario were related
to a charge-ordering process in the spirit of formation of a "disordered
Wigner crystal or, perhaps more appropriately, a "Wigner-glass"
\cite{39,40,41,42}. The visibility of shoulders in \cite{36} is exponentially
reduced with the disorder-strength W. Consequently, this feature has been
resolved in the simulations only for W \textit{much} smaller than the Coulomb
interaction amplitude E$_{\text{C}}$ \cite{38,43}. By contrast, the ratio
W/E$_{\text{C}}$ in Anderson-insulators with Gaussian disorder is typically
larger than unity \cite{24}. In the sample shown in Fig.1 for example, this
ratio is estimated to be $\approx$6, which, according to numerical simulations
\cite{36,38}, seems unfavorable for a charge-ordering mechanism to be
effective. To account for a "shouldered-MD" (SMD), it may perhaps be necessary
to take additional factors into account than included in the considerations
used for the classical Coulomb-glass.

We present in this work extensive data pertaining to this new phenomenon and
offer a heuristic interpretation for the main effects that characterize it.
Some puzzling issues that need further experimental and theoretical
elucidation are pointed out.

\section{Experimental}

\subsection{Sample preparation}

The samples used in this study were amorphous indium-oxide (In$_{\text{x}}$O)
films made by e-gun evaporation of 99.999\% pure In$_{\text{2}}$%
O$_{\text{3-x}}$ onto room-temperature Si-wafers in a partial pressure of
1.3x10$^{\text{-4}}$mBar of O$_{\text{2}}$ and a rate of 0.3$\pm$0.1\AA /s.
Under these conditions the carrier-concentration \textit{N} of the samples,
measured by the Hall-Effect at room-temperatures, was \textit{N}$\approx
$(1$\pm$0.1)x10$^{\text{19}}$cm$^{\text{-3}}$. Using free-electron formula,
this carrier-concentration is associated with $\partial$n/$\partial\mu\approx
$10$^{\text{32}}$erg$^{\text{-1}}$cm$^{\text{-3}}$. The Si wafers (boron-doped
with bulk resistivity $\rho\leq$2x10$^{\text{-3}}\Omega$cm) were employed as
the gate-electrode in the field-effect experiments. A thermally-grown
SiO$_{\text{2}}$ layer, 2$\mu$m thick, served as the spacer between the sample
and the conducting Si:B substrate. The screening length of the material
$\lambda$ $\approx$($\pi$e$^{\text{2}}\partial$n/$\partial\mu$)$^{\text{-1/2}%
}$ is $\approx$2nm and therefore the voltage that actually affects the sample
is $\approx$10$^{\text{-3}}$ of V$_{\text{g}}$, the voltage applied between
the sample and the gate. At room-temperature the sheet resistance of the
samples used here ranged between 22.5k$\Omega$ to 25k$\Omega.$

Films thickness was measured in-situ by a quartz-crystal monitor calibrated
against X-ray reflectometry. Samples geometry was defined by the use of a
stainless-steel mask during deposition into rectangular strips 0.9$\pm$0.1mm
long and 1$\pm$0.1mm wide. Four different batches were made in this study and
here we report detailed results of three different samples from one of these
batches with a thickness of 20$\pm$5nm. The dimensionless parameter
k$_{\text{F}}\ell$ for these samples is 0.27-0.29, just on the insulating side
of the critical value for this material (0.31$\pm$0.1 \cite{28}).

\subsection{Measurement techniques}

Conductivity of the samples was measured using a two-terminal ac technique
employing a 1211 ITHACO current preamplifier and a PAR 124A lock-in amplifier
using frequencies of 64-71Hz depending on the RC of the sample-gate structure.
R is the source-drain resistance and C is the capacitance between the sample
and the gate. C in our samples was typically $\cong$10$^{\text{{\small -10}}}%
$F and R for the samples studied in this work ranged between 1.5-2.5M$\Omega$
$at$ 4.11K. Except when otherwise noted, the ac voltage bias in conductivity
measurements was small enough to ensure near-ohmic conditions. Except where
otherwise noted, measurements were performed with the samples immersed in
liquid helium at T$\approx$4.11K held by a 100 liters storage-dewar. This
allowed up to two months measurements on a given sample while keeping it cold
and in the dark. These conditions are essential for measurements where
extended times of relaxation processes are required at a constant temperature.
Fuller details of the field-effect measurements setup, samples configuration
and characterization are described in \cite{29}. An up to date list of the
studied Anderson-insulators that exhibit a memory-dip along with a description
of their systematic dependence on the carrier-concentration of the material
are given in \cite{19}.

\section{Results and Discussion}

\subsection{The basic facts associated with the SMD}

The SMD state is sensitive to the conditions under which it is measured in the
field-effect experiment. In particular, sweeping V$_{\text{g}}$ over a range
$\vert$%
$\Delta$V$_{\text{g}}$%
$\vert$
that exceeds the typical width $\Gamma_{\text{S}}$ of the shoulder results in
a shoulderless MD when a V$_{\text{g}}$-sweep is taken again along the
original interval and polarity. The protocol used throughout this work to
achieve this state (labeled as "reference") involves sweeping V$_{\text{g}}$
from 0V to +70V and back to 0V then the sample is relaxed at this position for
20 minutes allowing the ZBA to reform before taking the next G(V$_{\text{g}}$)
scan. An example is illustrated in Fig.4:%

\begin{figure}[ptb]%
\centering
\includegraphics[
height=2.4785in,
width=3.3399in
]%
{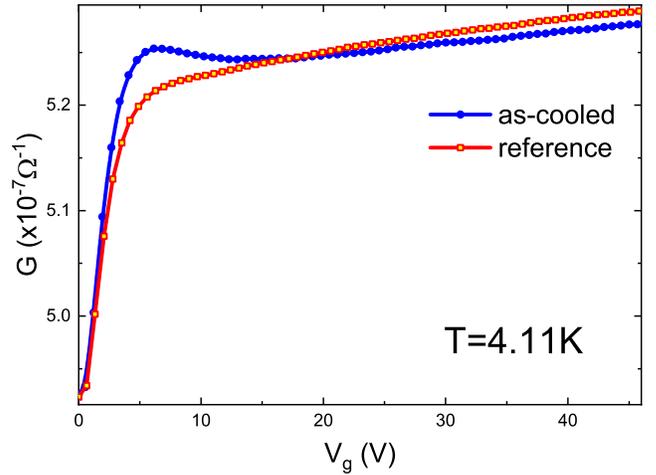}%
\caption{Field-effect scans comparing the reference, shoulderless state (see
text) with the G(V$_{\text{g}}$) of the sample 20 minutes after being
quench-cooled from room temperature to the bath-temperature. The two plots
were taken with the same scan-rate $\partial$V$_{\text{g}}$/$\partial$t of
1.2V/s.}%
\end{figure}

However, this shoulderless G(V$_{\text{g}}$) turns out to be metastable; the
SMD state of the sample reappears if one let the sample equilibrates under the
initial V$_{\text{g}}$ for long enough time before taking a new sweep. The
recovery of the SMD is illustrated in Fig.5 and Fig.6 for the right-hand and
left-hand side G(V$_{\text{g}}$) scans respectively.%

\begin{figure}[ptb]%
\centering
\includegraphics[
height=2.4215in,
width=3.3399in
]%
{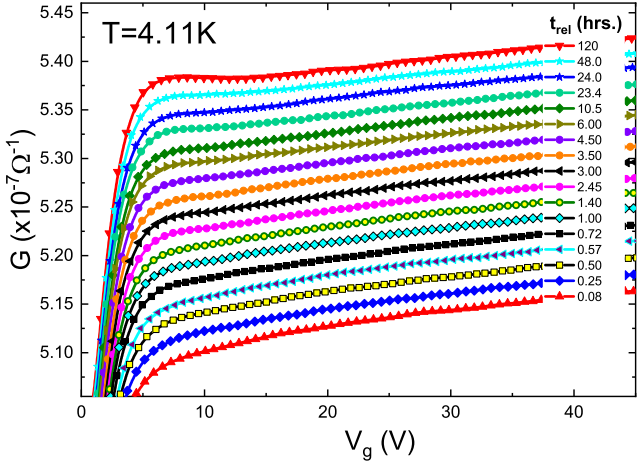}%
\caption{The recovery of the SMD-state with relaxation time t$_{\text{rel}}$
starting from a "reference" state at t=0.08 hours. Data are shown here for the
right-hand side of the field-effect. Before taking each plot, a new reference
state was prepared by sweeping the gate voltage to +70 V and back to 0V. The
plots are displaced along the ordinate for clarity and each was taken with the
same sweep-rate of dV$_{\text{g}}$/dt=1.2V/s.}%
\end{figure}
%

\begin{figure}[ptb]%
\centering
\includegraphics[
height=2.4128in,
width=3.3399in
]%
{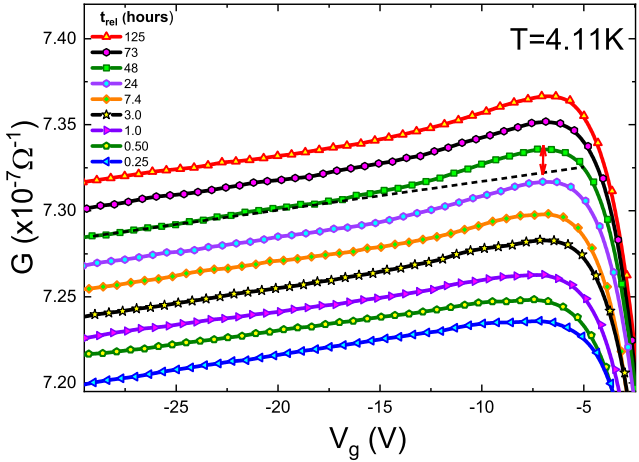}%
\caption{The recovery of the SMD-state as function of the relaxation time
t$_{\text{rel}}$ starting from a reference state at t=0.25 hours. Data are
shown for the left-hand side of the field-effect. Before taking each plot, a
new reference state was prepared by sweeping the gate voltage to +70 V and
back to 0V. The plots are displaced along the ordinate for clarity. All plots
were taken with the same sweep-rate of dV$_{\text{g}}$/dt=1.2V/s. The dashed
line and the red arrow delineate the way we define the magnitude of the
shoulder at a given t$_{\text{rel}}$ (see text and Fig.7).}%
\end{figure}
These time-dependent data imply that the SMD state has a lower energy than the
reference state.

The dynamics associated with the SMD `rejuvenation' process described by the
data in Figs. 5 and 6 differs from that exhibited by that of the
electron-glass relaxation-process monitored through the conductance G(t)
which, characteristically, is logarithmic \cite{27}. The two types of
relaxations are compared in Fig.7 for one of the samples:%

\begin{figure}[ptb]%
\centering
\includegraphics[
height=3.3062in,
width=3.3399in
]%
{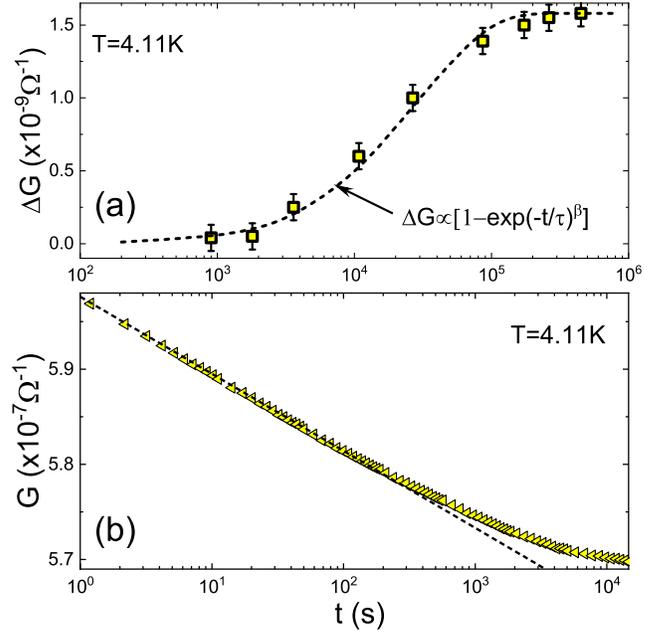}%
\caption{(a): The recovery of the shoulder with time characterized by $\Delta
$G defined by the construction shown in Fig.6 as a red arrow. The dashed line
in the figure is a fit to the stretched-exponential function with $\beta$=0.78
and $\tau$=23,400s. (b): The relaxation of this sample conductance monitored
by recording G(t) at V$_{\text{g}}$=45V after changing the gate-voltage from
the equilibrium state at 0V to 45V. The dashed line delineates the logarithmic
relaxation law. Note the two decades difference in time scales between the two
plots.}%
\end{figure}

Note first that the SMD evolution persists and is followed for two more
decades in time than the conductance relaxation. The main part of the change
in G(t) is over while that of $\Delta$G is still building up. Secondly,
although the two processes are probably related (and begin and end essentially
at the same times), the functional time dependence of $\Delta$G(t) being
stretched exponential with exponent $\beta$=0.78, suggests a process dominated
by a relatively narrow rate-distribution \cite{44}. This point will be
clarified below.

The disparity in the dynamics of the two types of slow relaxations is also
manifested in the dependence of the SMD magnitude on the sweep-rate. This is
shown in Fig.8 for another sample in the batch studied.%

\begin{figure}[ptb]%
\centering
\includegraphics[
height=3.8493in,
width=3.3399in
]%
{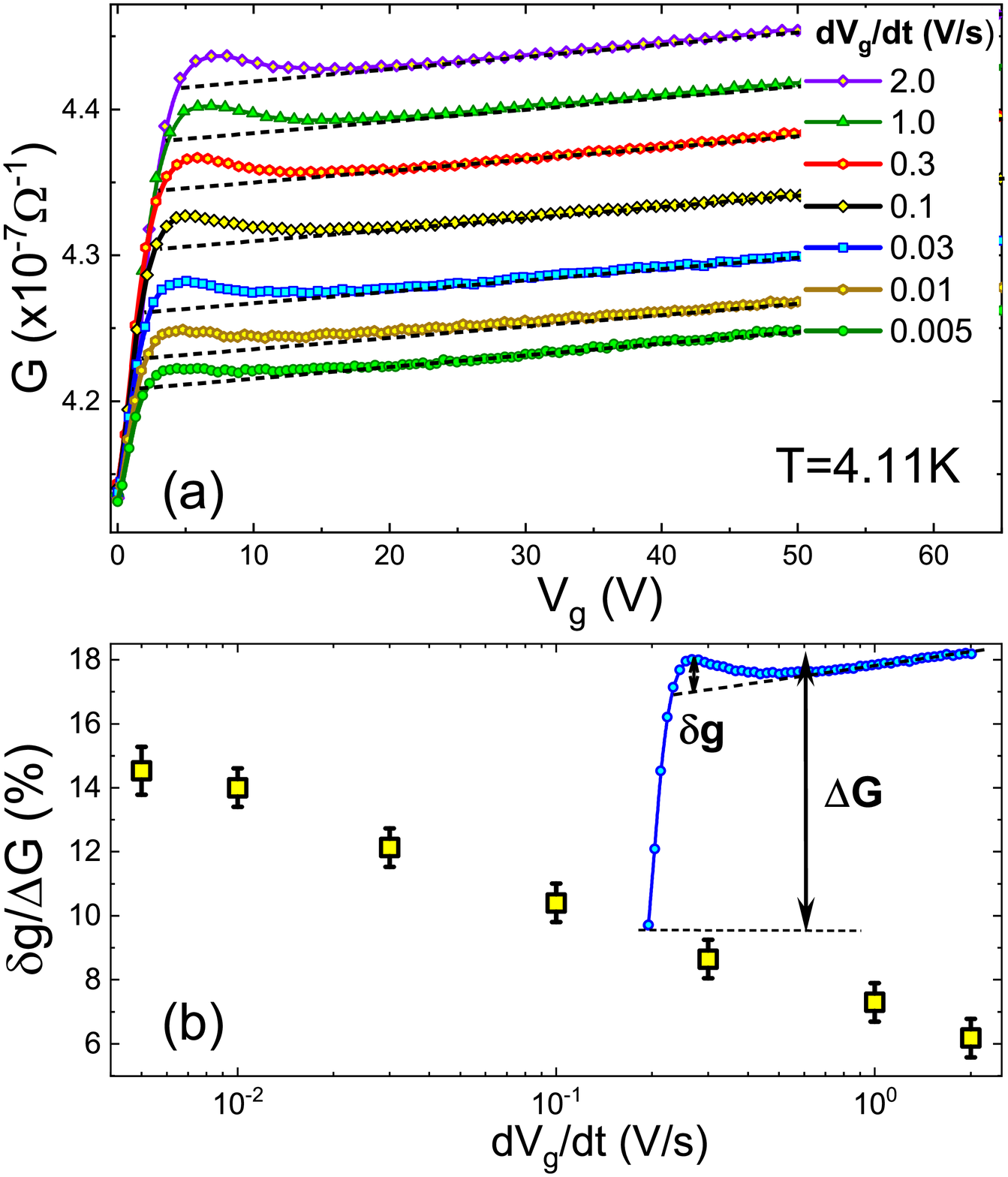}%
\caption{(a): Dependence of the SMD plots on the gate-voltage sweep-rate
dV$_{\text{g}}$/dt. (b): The magnitude of the shoulder relative to that of the
MD, defined as shown in the inset, as function of the sweep-rate
dV$_{\text{g}}$/dt.}%
\end{figure}
Note that $\Delta$G, the depth of the ZBA in these plots is changing
substantially with the sweep-rate (Fig.8a) owing to the relative fast
relaxation dynamics of these samples. The logarithmic dependence of $\Delta$G
on the sweep-rate is a common feature in electron-glasses \cite{45}. This
dependence carries-over to the ratio $\delta$G/$\Delta$G ( where $\delta$G is
the shoulder height, see: Fig.8b) because of the much weaker change of
$\delta$G with the sweep-rate.

\subsection{The shoulder origin - a heuristic interpretation}

Based on the data presented above, it is conjectured that the shoulders
reflect an increase in the density of states at the edges of the MD in the
same vein that the depression of G(V$_{\text{g}}$) signifies a depletion of
charge at the chemical potential set by v$_{\text{g}}^{\text{eq.}}$, the
equilibrium gate-voltage. The MD has been related to the soft gap at $\rho
$($\varepsilon$) resulting from the long-range Coulomb interaction
\cite{12,13,14,17,18,19,45}.

On its own, the interaction favors a correlated spatial distribution of the
states expelled from the low energy of $\rho$($\varepsilon$). Therefore, a
lower energy configuration should be attained when the distribution of the
nearest-neighbor distances is peaked at a value of the order \textit{N}%
$^{\text{-1/3}}$ where \textit{N} is the system carrier-concentration. This in
turn would show up as a peak in G(V$_{\text{g}}$) at E$_{\text{C}}\approx
$e$^{\text{2}}\mathit{N}^{\text{1/3}}$/$\kappa$ where $\kappa$ is the
effective dielectric constant of the medium. The shoulder in our samples is
peaked at E$_{\text{C}}\approx$7~meV and it tapers-off towards $\approx
$40-50~meV (Fig.1). For the batch used here, \textit{N}$\approx$%
10$^{\text{19}}$cm$^{\text{-3}}$, the observed peak of the shoulder is
consistent with E$_{\text{C}}\approx$e$^{\text{2}}\mathit{N}^{\text{1/3}}%
$/$\kappa$ if $\kappa\approx$45. The bare dielectric constant of the material
is $\kappa\approx$10 but the polarization of localized states may enhance it
to be consistent with this scenario.

\ To further asses the plausibility of this picture, we need to consider the
role of disorder. The precondition for the MD to be observable in the
field-effect experiment is a strong enough disorder \cite{19}. Explicitly, the
disorder has to be strong enough to Anderson-localize the system \textit{and}
to render its relaxation slower \cite{19} than the sweep-rate of V$_{\text{g}%
}$. On the other hand, too strong a disorder would defeat the
electron-electron tendency to form an ordered structure and randomize the
states spatial-distribution thus suppressing the shoulder. Intuitively, out of
the systems that exhibit MD, the best candidate to show shoulders is where the
disorder to interaction ratio is small. This favors systems with low
carrier-concentration \textit{N} because the disorder necessary to
Anderson-localize a system increases with its Fermi-energy E$_{\text{F}%
}\propto$\textit{N}$^{\text{\textit{2/3}}}$ while the interaction scales with
\textit{N}$^{\text{\textit{1/3}}}$. Note that the SMD discussed here has the
lowest \textit{N} among all the electron-glasses systems\textit{ }yet reported
\cite{19,46}. This may be in line with our charge-ordering scenario in the
sense that the present system has more favorable parameters than previously
studied electron-glasses to show shoulders in field-effect measurements.

But how favorable are the actual system parameters? The magnitude of the
interaction identified above with the position of the shoulder was
E$_{\text{C}}\approx$7~meV. The relevant \cite{47} disorder W may be estimated
from the width of the shoulder (Fig.1) yielding W/E$_{\text{C}}\approx$6. The
disorder to interaction ratio typical for a degenerate semiconductor ranges
between $\approx$3 to $\approx$100 \cite{24}. As anticipated, the current
system is at the "favorable" limit in terms of parameters. Yet, this ratio is
far from the W/E$_{\text{C}}\ll$1 limit where shoulders are observed in
classical simulations \cite{36,37,38}. Hopefully, the results reported here
may assist in identifying the missing ingredients from these simulations.

The presence of disorder is also essential for understanding the
time-dependent processes depicted in Figs.5 and 6; Sweeping the gate voltage
by a
$\vert$%
$\Delta$V$_{\text{g}}$%
$\vert$%
$>$%
$\Gamma_{\text{S}}$ (where $\Gamma_{\text{S}}$ is the shoulder-width) results
in a shoulderless state (Fig.4). However, given time, the shoulder is
sluggishly recovered and G(V$_{\text{g}}$) eventually exhibits the SMD
behavior. Presumably, the metastable MD-state is made up of charges trapped by
disorder over the voltage-range swept by the gate. Some of the charges that
were driven by the gate to occupy higher energy-states are trapped by deep
wells of the potential while the gate travels back to v$_{\text{g}%
}^{\text{eq.}}$. Charges occupying states with shallow wells of the potential
will escape these states during the time the gate is set back in v$_{\text{g}%
}^{\text{eq.}}$ to rebuild the MD for the next G(V$_{\text{g}}$) run. This
explains both, the shifting of charge away from the shoulder towards higher
energies, and the relative absence of the fast relaxation rates in the
relaxation associated with the shoulder recovery.

The proposed picture for creating the shoulderless state is analogous to the
process by which debris is washed ashore by sea-waves and is held on the beach
slope by friction when the wave pulls back seaward.

Despite the apparent difference in their dynamics, it is clear that the
shoulder and the memory-dip are two parts of the same phenomenon; when the
equilibrium voltage is moved, so is the position of the shoulder. An example,
illustrating the `two-dip-experiment' \cite{46} for the current batch of
In$_{\text{x}}$O, is shown in Fig.9:%

\begin{figure}[ptb]%
\centering
\includegraphics[
height=2.4137in,
width=3.3399in
]%
{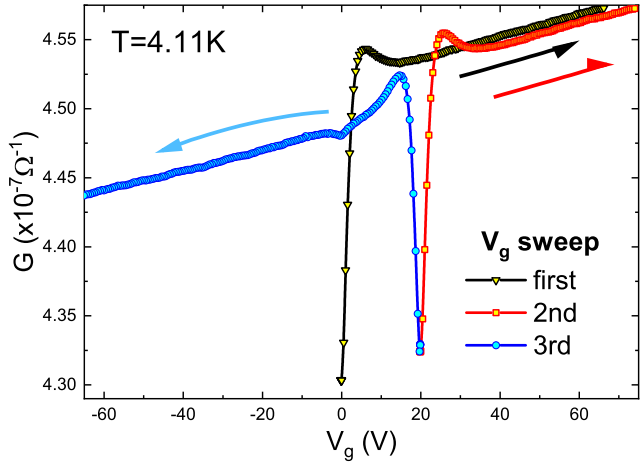}%
\caption{G(V$_{\text{g}}$) plots taken consecutively; First, V$_{\text{g}}$
was swept from the original equilibrium v$_{\text{g}}^{\text{eq.}}$=0V to +65V
(triangles). This is followed by sweeping V$_{\text{g}}$ to +20V and parked
there for 20 minutes. Next, V$_{\text{g}}$ was swept to +80V (squares).
V$_{\text{g}}$ was then moved to +20V and parked there for 20 minutes to
relax. Finally, V$_{\text{g}}$ was swept to -65V (circles). Note that the
latter G(V$_{\text{g}}$) plot was taken less than an hour after the first
sweep and the magnitude of the MD at v$_{\text{g}}$=0V has shrank by a factor
of $\approx$60 relative to the original value. All plots were taken with the
same dV$_{\text{g}}$/dt=0.5V/s.}%
\end{figure}

The feature that strikes the eye in Fig.9 is the barely discernible MD
produced in the third sweep at the original v$_{\text{g}}^{\text{eq.}}$=0V.
The magnitude of this MD is smaller by almost two orders of magnitude relative
to either of the two "fresh" MD's. This is due to the relatively fast dynamics
of this weak-disorder batch \cite{19,46,48}. A closer look at the figure
reveals a new aspect of the phenomenon; establishing a new v$_{\text{g}%
}^{\text{eq.}}$(at +20V) yields at the edge of the MD a prominent shoulder
despite being in voltage-range covered by V$_{\text{g}}$ in a previous sweep.
In a way, this is similar to that the shoulders appear along with the MD upon
the first cooldown from room-temperatures. Both the MD and the shoulder appear
in this case after a brief relaxation period (typically 20 minutes), no
days-long waiting-time is required. In fact, as will be shown next, once a new
equilibrium conditions are set, the full SMD appears in the G(V$_{\text{g}}$) plot.

\subsection{Shoulder destruction and rejuvenation}

As shown above, taking a G(V$_{\text{g}}$) measurement over a V$_{\text{g}}%
$-range larger than $\Gamma_{\text{S}}$, eliminates the shoulder from
showing-up in the next scan. A prolonged relaxation of the sample is required
for its recovery. However, there are several ways to restore the SMD without a
long time-delay. Two of these were mentioned in the paragraph above. Three
other schemes effective in rejuvenating the shoulder are described next.

The SMD's resulting from applying the first two are illustrated in Fig.10.%

\begin{figure}[ptb]%
\centering
\includegraphics[
height=2.4422in,
width=3.3399in
]%
{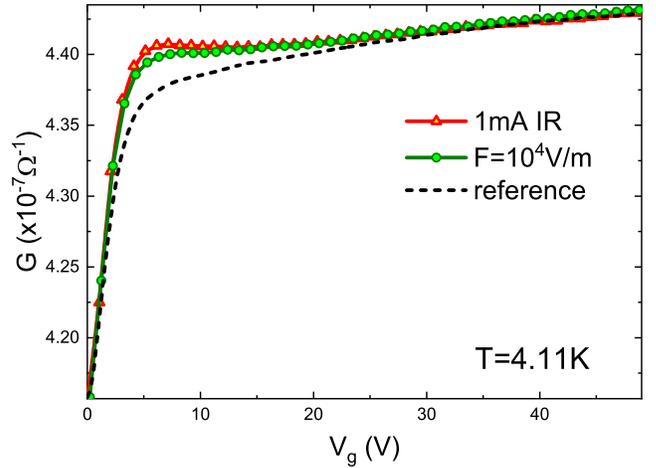}%
\caption{Field-effect plots illustrating two examples of the shoulder
reappearance following a quench from a nonequilibrium, higher-energy state:
Stressing the sample with a non-ohmic field F=10$^{\text{4}}$V/m held on for
20 minutes (circles), and exposing the sample to 1mA of the IR source for 3s
(triangles). In both cases the shown G(V$_{\text{g}}$) plots are taken after
letting the sample relax for 20 minutes since the respective quench. These are
contrasted with the reference plot produced after a scan to +70V. All three
plots were measured in the linear-response and with the same dV$_{\text{g}}%
$/dt=1V/s.}%
\end{figure}

Both schemes produce shoulders that exhibit higher visibility than would be
expected by allowing the system to relax at v$_{\text{g}}^{\text{eq.}}$ for
the time required for completing the scheme (compare with Fig. 5).

The third scheme is the easiest to implement and it turns out to be the most
effective in pulling back the shoulder from the reference state. In this case,
V$_{\text{g}}$ is first wept to V$_{\text{g}}$=-%
$\vert$%
v$_{\text{g}}^{\text{bias}}$%
$\vert$%
$_{\text{1}}$ staying there for $\approx$2 seconds and then set back at
V$_{\text{g}}$=0V for 20 minutes relaxation before recording a G(V$_{\text{g}%
}$) plot from 0V to +60V. As noted before, this trip erases the shoulder. This
"v$_{\text{g}}^{\text{bias}}$" protocol is repeated for different (negative)
bias-values thus generating plots as function of -%
$\vert$%
v$_{\text{g}}^{\text{bias}}$%
$\vert$%
. In each such plot the shoulder is "rejuvenated" by pre-biasing the sample
with a finite -%
$\vert$%
v$_{\text{g}}^{\text{bias}}$. These bias values were taken at random order to
verify that `history' does not play a role. A series of such plots with
$\vert$%
v$_{\text{g}}^{\text{bias}}$%
$\vert$
values ranging from 5V to 80V is shown in Fig.11a.%

\begin{figure}[ptb]%
\centering
\includegraphics[
height=4.1796in,
width=3.3399in
]%
{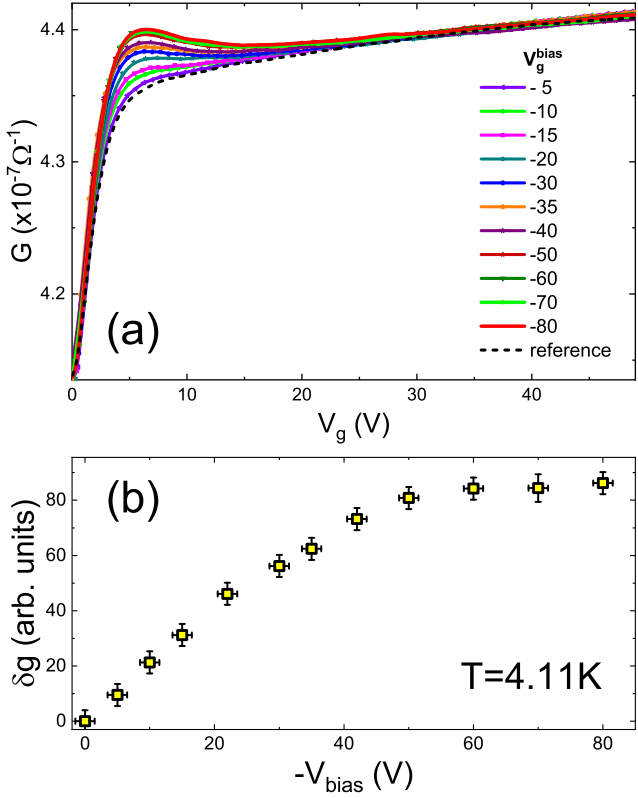}%
\caption{The reappearance of the shoulder in the G(V$_{\text{g}}$) plots
recorded by the "v$_{\text{g}}^{\text{bias}}$ protocol" described in the text.
These are labeled by their pre-scan (negative) bias (a). The plots were taken
consecutively using the same sweep rate dV$_{\text{g}}$/dt=1V/s. (b) The
magnitude of the shoulder vs. the pre-bias voltage. This magnitude is taken as
the distance from the peak of the shoulder at 6.8V to the value of the
reference-curve at this voltage{\protect\small .}}%
\end{figure}

Note (Fig.11b) that the relative magnitude of the shoulder $\delta$G increases
with
$\vert$%
v$_{\text{g}}^{\text{bias}}$%
$\vert$
and tends to saturate for v$_{\text{g}}^{\text{bias}}\simeq$-50V. The range
over which the gate-voltage bias affects $\delta$G is similar to the typical
width of the shoulder $\Gamma_{\text{S}}$ that, in turn, is related to the
disorder associated with the phenomenon W. It is emphasized that these data
are generated using random values for
$\vert$%
v$_{\text{g}}^{\text{bias}}$%
$\vert$
in the consecutive measurements. Therefore the correlation between
$\vert$%
v$_{\text{g}}^{\text{bias}}$%
$\vert$
and $\delta$G is meaningful and suggestive. Actually, these results lead us to
recognize the common element in the four different schemes that produces a SMD
from a "reference" state: \textit{They all involve a quench from a state where
the charge-carriers are endowed with excess energy }E$_{\text{excess}}%
$\textit{ relative to the equilibrium state}. These schemes differ by how
E$_{\text{excess}}$ is structured (distributed in energy) relative to W, which
presumably determines its efficiency in pulling-up the shoulder. When the
excess-energy of charge-carriers exceeds W, they are no longer trapped by
potential wells of the disorder. Rather, their spatial distribution is
controlled by the Coulomb repulsion forcing them apart. In this stage the
system is in a fluid, spatially-correlated state that, following a quench and
brief relaxation, is well primed to form SMD.

The efficiency of a scheme to produce SMD is not necessarily reflected in the
conductance enhancement accompanying the process. For example, stressing the
sample with a large longitudinal field enhances the sample conductance by a
factor of $\approx$2 while the shoulder that results from this scheme is
visibly weaker (Fig.10) than that obtained by the brief IR illumination that
had a much smaller impact on the conductance (see Fig.12a,b).%

\begin{figure}[ptb]%
\centering
\includegraphics[
height=4.2298in,
width=3.3399in
]%
{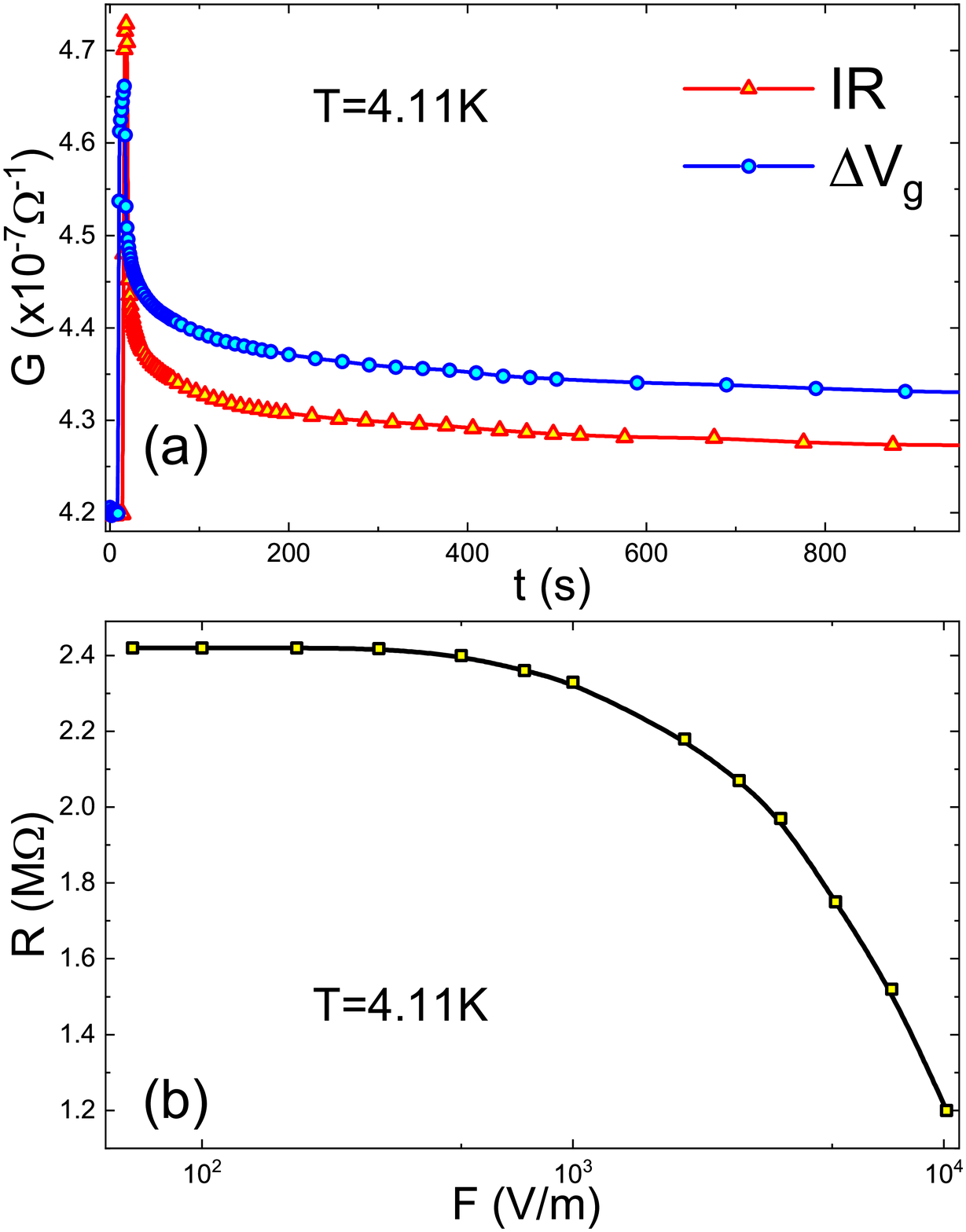}%
\caption{The effects produced on the sample by a brief IR illumination and a
large applied electric-field. (a) The conductance increase and subsequent
relaxation in response to a 3s exposure to the 1mA IR source. This is compared
to the similar effect produced by a sudden change of the gate-voltage from 0V
to 45V (a 5 seconds trip-time). (b) The sample resistance as function of the
applied electric field.}%
\end{figure}

\subsubsection{Temperature dependence}

Temperature has a marked effect on the electron-glass features. Both the width
of the memory-dip and its relative magnitude are systematically affected by
raising the temperature \cite{49}. Here it is shown that the shoulder follows
the same trend; its peak is slightly shifted towards a higher V$_{\text{g}}$
upon increasing the bath temperature, and its magnitude is exponentially
suppressed (Fig.13b). This temperature dependence was obtained before in the
electron-glass phase of In$_{\text{2}}$O$_{\text{3-x}}$ films \cite{49}.%

\begin{figure}[ptb]%
\centering
\includegraphics[
height=4.2021in,
width=3.3399in
]%
{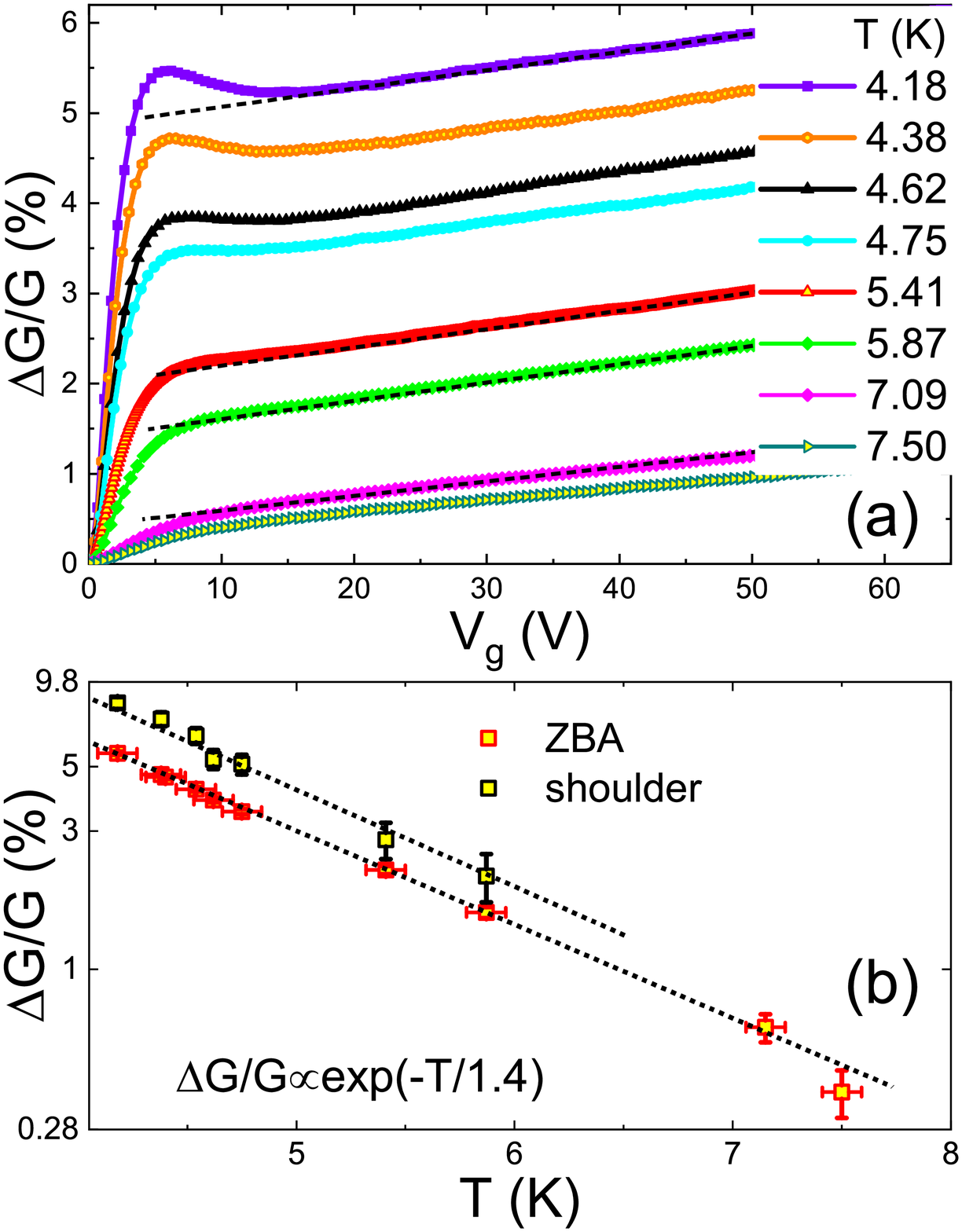}%
\caption{(a): The dependence of the G(V$_{\text{g}}$) plot on temperature for
a typical SMD sample. Each plot was taken after first sweeping the gate to
-70V and back to 0V to allow shoulder rejuvenation (see text) and relaxation.
Then the bath temperature was changed and the same protocol repeated with
V$_{\text{g}}$/dt=0.5V/s. (b): The relative magnitude of the ZBA and the
shoulder extracted from the data in (a). $\Delta$G/G for the ZBA is the value
of G(V$_{\text{g}}$) at V$_{\text{g}}$=7V. $\Delta$G/G values for the shoulder
are taken as in Fig. 6 relative to the dashed lines in (a). Dotted-lines fit
exp[-T/1.4] (see text). Note that both axes in plot (b) use logarithmic
scale.}%
\end{figure}

Apart from reaffirming that the shoulder is an integral part of the
memory-dip, it illustrates how sensitive is its visibility to the addition of
high-energy components.

\subsubsection{Nonequilibrium steady-state under IR illumination}

Even more striking is the effect of a continuous IR illumination which
suppresses the shoulder despite the extremely weak power-load on the sample (Fig.14):%

\begin{figure}[ptb]%
\centering
\includegraphics[
height=3.5976in,
width=3.3399in
]%
{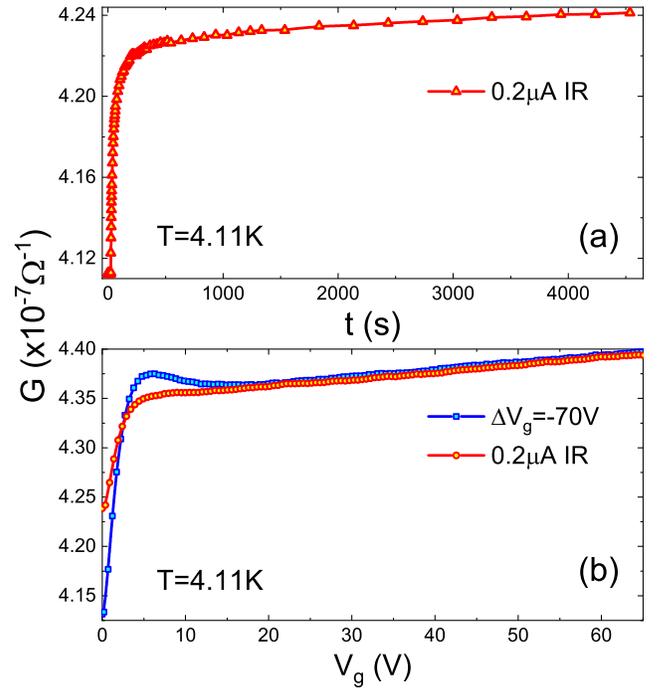}%
\caption{ Sample conductance versus time while being exposed it to a IR
radiation with a constant power of 2\textperiodcentered10$^{\text{-7}}$Watt.
Note that G(t) tends to reach a saturated value for t$\gtrsim$1 hour (b): The
sample G(V$_{\text{g}}$) plot in the dark (squares) compared with the
G(V$_{\text{g}}$) plot taken under the constant IR illumination after the
conductance reached a "saturated" conditions (explicitly, the change with time
for the duration of the G(V$_{\text{g}}$) plot is negligible). Both curves
were taken after first sweeping V$_{\text{g}}$ to -70V, then relaxing the
sample for 20 minutes, and using the same sweep-rate dV$_{\text{g}}%
$/dt=0.5V/s.}%
\end{figure}
Under a continuous IR illumination the sample SMD magnitude was reduced to
half its dark-value while the conductance increased by only $\simeq$2.8\%
(see, Fig.14a). Based on the conductance versus temperature of this sample
shown in Fig.15 this is the equivalent of raising the temperature by $\approx
$20mK. To achieve a similar reduction of the SMD magnitude of the same sample,
the bath temperature must be raised to T$\simeq$5.5K (see Fig.13) which
increases the conductance by $\simeq$300\%.

The difference between these two protocols is in how the energy added to the
system is distributed among the available degrees of freedom. Under the IR
illumination, the system is populated with an excess of high-energy phonons.
These are generated by a cascade process: Electrons are excited to
high-energy, then relax by optical-phonons emission \cite{50}. This
nonequilibrium steady-state has much fewer low-energy phonons than when the
sample is heated-up to $\approx$5.5K. The density of high-energy phonons is
also increased by raising the bath temperature, but most of the absorbed
energy is spent on enhancing the conductance. The indication is that the
SMD-state is sensitive to the presence of high-energy phonons. This may hint
on an underlying mechanism that counteracts the disorder in the competition
with interactions. High-energy phonons is a potent source of dephasing which
raises the possibility that quantum effects play a role. Effects that hinge on
quantum-coherence and may be relevant to the problem at hand are those that
promote spatially-extended wavefunctions like resonances \cite{51,52,53} and
other wavefunction forms resulting from hybridization. The existence of
extended wavefunctions naturally mimics a weaker disorder. There is already
evidence for quantum coherence in this system \cite{50, 55} so there is reason
to further explore this direction.%

\begin{figure}[ptb]%
\centering
\includegraphics[
height=2.9758in,
width=3.3399in
]%
{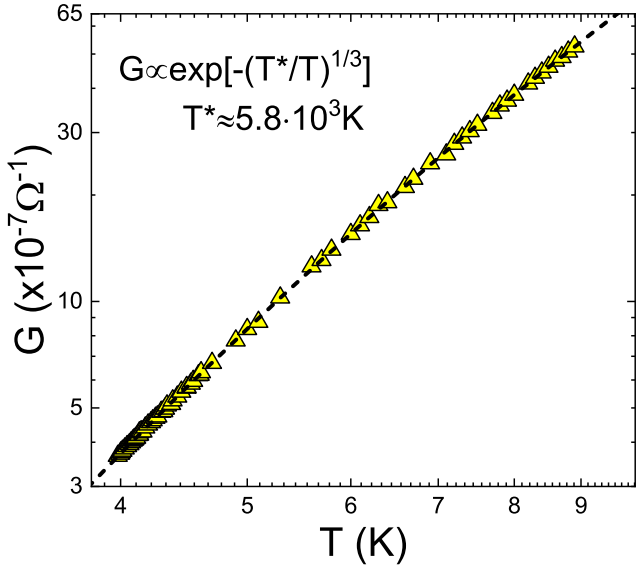}%
\caption{The temperature dependence (on a double-log scale) of the conductance
for the sample used in figures 8-14 inclusive}%
\end{figure}

\subsection{Discussion}

There are several issues related to the appearance of the shoulders that need
to be further clarified. First is the absence of this feature in other
versions of In$_{\text{x}}$O. In contrast with the samples reported here, all
studied Anderson-insulating In$_{\text{x}}$O samples with \textit{N }ranging
from\textit{ }$\approx$4x10$^{\text{19}}$cm$^{\text{-3}}$ to $\approx
$8x10$^{\text{21}}$cm$^{\text{-3}}$ exhibit electron-glass characteristics
with MD magnitude and width varying systematically with \textit{N } but
without shoulders \cite{19,27,48}. It seems therefore that, for In$_{\text{x}%
}$O, as \textit{N }increases from $\approx$1x10$^{\text{19}}$cm$^{\text{-3}}$
to $\approx$4x10$^{\text{19}}$cm$^{\text{-3}}$, the structure manifested by
the shoulders melts-away. In terms of In/O ratio, which is the way \textit{N}
is determined in the compound, this range is merely $\approx$2\% wide
\cite{28}. It would be interesting to fine-tune the system \textit{N} and
track the evolution of the phenomenon across this region. This is a much
harder undertaking than it may seem and will require establishing a new level
of material deposition and post-treatment control. Nonetheless, on the basis
of what is already established, this SMD$\rightarrow$MD transition is
surprisingly sharp.

It appears that one has to decide between two possibilities: Either the
transition is a real phase-change, or it is a crossover to a regime with ultra
slow dynamics. Explicitly, as \textit{N} increases, the shoulders-dynamics
becomes too slow to be resolved on a realistic time scale. This dilemma is
similar to the quandary that has haunted the glass community \cite{54}, and
perhaps it may take as long to reach a conclusion. In the present case the
possibility of a diverging relaxation time has empirical support; The
relaxation time of In$_{\text{x}}$O electron-glasses with \textit{N}$\approx
$4x10$^{\text{19}}$cm$^{\text{-3}}$ to \textit{N}$\approx$10$^{\text{20}}%
$cm$^{\text{-3}}$ was shown to increase dramatically with \textit{N}
\cite{45,46}, and as seen here. the dynamics associated with the shoulders may
actually be slower.

Yet, the possibility that the SMD to MD changeover is a true phase-transition
cannot be ruled-out either. If the MD$\rightarrow$SMD transition is indeed a
change between two distinct phases what are their nature? The SMD state,
characterized by a shoulder in the G(V$_{\text{g}}$) plot has been interpreted
here as a preponderance of charge at energy of the order E$_{\text{C}}$
relative to the chemical-potential. It is illuminating to contrast this
picture with the spatial-order of the amorphous structure manifested in the
electron-diffraction patterns (Fig.2, and Fig.3 above). The shoulders and the
diffraction-pattern represent two amorphous structures; one composed of ions,
and the other of electronic-charges. The first ring in the former is related
to the nearest-neighbor distance between ions which is the analog of the
shoulder position in energy. Both the diffraction-ring and the shoulder have a
width that is a measure of the disorder associated with the respective
phenomenon. The diffraction-ring is relatively much sharper than the
shoulder-width. This is due to the difference in the two-particle potential
that constrain spatial-freedom of the constituents; Lennard-Jones versus
Coulomb-interaction for the ions and charge-carriers respectively. For
energies outside the Coulomb gap, this description depicts the SMD-state as a
"frozen-liquid" while the MD-state, lacking `medium-range-order', is a
"frozen-gas". The experiments that were performed up to and including this
work indicate that there is a transition between these states controlled by
the carrier-concentration \textit{N}. This was ascribed to the competition
between disorder and interaction that is tilted in favor of the former when
\textit{N} increases. To further test this picture one needs to change
disorder, or the interaction, independently which is a challenging task.

A related issue is the use of lower carrier-concentrations than \textit{N}%
$\approx$10$^{\text{19}}$cm$^{\text{-3}}.$ As mentioned above, in the
Anderson-localized regime (a precondition to show electron-glass features
\cite{19}), both disorder W and the Coulomb-interaction E$_{\text{C}}$
increase with \textit{N} but W$\propto$\textit{N}$^{\text{2/3}}$ while
E$_{\text{C}}\propto$\textit{N}$^{\text{1/3}}$ and therefore a lower
disorder-interaction ratio is achievable by reducing \textit{N}. This however
has a limited scope; the ratio is a weak function of \textit{N} while the
dynamics, mainly controlled by W \cite{19,46}, depends on it exponentially.
When W decreases the relaxation will quickly become too fast for the
experimental $\partial$V$_{\text{g}}$/$\partial$t to expose a memory-dip in
the field-effect experiments. To our knowledge, there is no Anderson-insulator
with \textit{N}$\lesssim$10$^{\text{18}}$cm$^{\text{-3 }}$that exhibits either
MD or SMD.

There are other factors that may be detrimental for charge-ordering besides
too strong disorder. Obviously, sample inhomogeneity beyond that inherent in
the sample disorder (estimated `globally', usually on the basis of measuring
the resistance) will smear-out the shoulders even when all other conditions
are favorable. In particular, an excess of randomly distributed deep
potential-wells, should be avoided. Generally, the detailed structure of the
background potential, not just its average amplitude, is relevant for the
charge-ordering scenario.

A subtle question that is yet to be answered is the effect of a superimposed
periodic potential. The effect of underlying crystallinity on the
disorder-interaction balance is a pertinent question that may be tested by
numerical simulations. However, numerical studies usually use a lattice-model
with half-filling, which probably introduces a spatial constraint that likely
is incommensurate with the optimal charge arrangement. It is not clear that it
may suffice to employ just a very small filling-factor in the simulation to
capture the physics of charge-ordering.

Attempts to answer some of these questions may benefit from further
experimental work. Systems that exhibit electron-glass features, such as GeTe,
Tl$_{\text{x}}$O, In$_{\text{2}}$O$_{\text{3-x}}$, GeSb$_{\text{2}}%
$Te$_{\text{5}}$, and Be \cite{19} do not show shoulders in their
G(V$_{\text{g}}$) plots. It would be useful to know whether this is due to
unfavorable W/E$_{\text{C}}$ or because of their crystallinity. Filling factor
in these compounds is naturally small but their relatively high
carrier-concentration may be unfavorable. It is worth trying to tweak this by
varying their composition. These systems should also be tested for
quantum-coherence as this may play a role as alluded to above. A methodical
study of these aspects may establish a new platform for the interplay between
disorder and interaction in Anderson insulators.

\begin{acknowledgments}
Discussions with Oded Agam, and Ady Vaknin are gratefully acknowledged.
\end{acknowledgments}


\begin{thebibliography}{99}
\bibitem {47}The disorder that is required to localize the system is related
to the competition with the system kinetic energy. It may therefore differ
from that associated with the competition with the Coulomb interaction.
\end{thebibliography}
\end{document}